\documentclass[a4paper,10pt]{article}
\usepackage[utf8]{inputenc}
\usepackage{geometry}
\usepackage{pgf,xcolor,tikz}
\title{Multi-class fundamental diagrams from the Prigogine-Herman-Boltzmann equation}
\author{A. R. M\'endez, W. Marques Junior, R. M. Velasco}

\begin{document}

\maketitle

\begin{abstract}
  Our aim in this paper is to establish a theoretical fundamental diagram for a multi-class traffic flow from a gas-kinetic-like
  traffic model. We start with a multi-class 
generalization of the Prigogine-Herman-Boltzmann equation to construct the fundamental relation for this system. We 
show that there exists a critical density which depends on the relative concentration of slow and fast users and describe a 
procedure to find the threshold value. Finally, our flow-density relation for a two-class mixture of vehicles is 
contrasted with empirical data in the literature.
\end{abstract}

\section{Introduction\label{intro}}
The fundamental diagram in traffic flow modeling is considered as the corner stone since the Greenshields \cite{Greenshields35} 
studies at the beginning of last century. The fundamental diagram relates the average speed and the density in an homogeneous and 
steady state, which is usually quoted as \textquotedblleft equilibrium\textquotedblright, even though it can be formulated also 
for a non-equilibrium state \cite{Kerner04}, \cite{Treiber10}.
 
There have been several approaches to describe and understand the traffic dynamics, however all of them start or end with results 
consistent with the fundamental diagram \cite{Kerner04}, \cite{Kerner09}, \cite{Treiber10}, \cite{Helbing01}. It should be 
noticed that those may come from an empirical data fit or have their origin on proposals which take into account the general 
characteristics associated with the behavior of vehicles in a traffic situation. The emergence  and evolution of fundamental 
diagrams study has been reviewed some years ago in reference \cite{Wageningen14}, where it is given a very good perspective 
about the status played by the fundamental diagram in traffic flow understanding. Nevertheless, to the author's knowledge, the 
construction of a theoretical multi-class fundamental relation based on a mesoscopic equation has never been addresed before 
in the literature.

The requirements which must be satisfied to give a sound relation for the fundamental diagram were synthesized in reference 
\cite{delCastillo95},\cite{delCastillo12} where it is said that (a) the speed values satisfy the condition  
$0\leq v\leq v_{max}$, (b) and the density $0 \leq \rho \leq \rho_{max}$. There are specific speed values 
(c) $v(\rho=0)=v_{max},~v(\rho_{max})=0$, (d) the flow is limited by $q(\rho=0)=0$ and $q(\rho_{max})=0$. Besides (e) the free 
flow speed is determined by $v_f=(dq/d\rho)_{\rho=0}$ and, the kinematic wave speed is given as $w=(dq/d\rho)_{\rho_{max}}$. 
Also (f) the fundamental diagram is strictly concave, so that $(d^2q/d\rho^2)<0$ for almost all $0\leq \rho \leq \rho_{max}$. 
Lastly (g) there exists a maximum flow $q_{max}=$  max $[q(\rho)]_{0<\rho <\rho_{max}}$ which assures that there are a critical 
value for the  density and speed such that $q_{max}=\rho_{crit}v_{crit}$. The critical density separates the fundamental diagram 
in two regions: (i) the free flow characterized by $\rho<\rho_{crit}$ with increasing flow for increasing density and, (ii) 
the congestion region with densities $\rho>\rho_{crit}$ and decreasing flow for increasing density. Most fundamental diagrams 
satisfy these conditions except the concavity property which do not seem to be necessary \cite{Zhang01}, instead a weaker 
condition is usually taken which assumes that the speed does not increase with the increasing density ${dv}/{d\rho} \leq 0$. 

Other aspect that must be considered in traffic modeling is the heterogeneity produced by the presence of several classes of 
vehicles and drivers on the road. Such heterogeneity has been identified as the cause of some phenomena, for example the 
capacity drop and the scattered behavior in the congested region, among others. This problem has been discussed in the 
literature in some papers \cite{Buison09}, \cite{Cho02}, \cite{Gupta07}, where some multi-class mesoscopic models were 
constructed in terms of different assumptions such as the three macroscopic variables \cite{Hoogendoorn00}. There has been 
some other trends to model heterogeneous traffic flow by means of a generalization of the well known LWR model \cite{Wong02}, 
all of them give some prescription to consider the interaction between classes \cite{Ngoduy10}, \cite{Ngoduy11}, 
\cite{vanWageningen13}, \cite{Zhang06}.  
    
In this paper we start with the kinetic equation proposed by Prigogine and Herman \cite{Prigogine71}, which resembles the 
Boltzmann kinetic equation to describe the behavior for a dilute gas and will be called as Prigogine-Herman-Boltzmann equation 
(PHB), to construct the fundamental diagram in an equilibrium state. The kinetic equation considers that the distribution 
function $f(x,c,t)$ tends to coincide with a speed desired distribution function $f^0(x,c,t)$  in a relaxation time 
$\tau(\rho)$. Besides there are interaction terms between vehicles through an overtaking probability $p(\rho)$ which 
depends on the density. When we consider just one class of drivers, the procedure to obtain the fundamental diagram follows 
similar steps as those followed in \cite{Prigogine71} and \cite{Iannini16}, where the equilibrium distribution function is 
obtained in terms of a desired one which can be chosen. Each selection drives to a different fundamental relation
which must be consistent with the general characteristics mentioned above. As a second goal in our paper, the two-classes fundamental 
diagram is also obtained and compared with empirical findings.  

In Section \ref{PHB} we obtain the distribution functions for the equilibrium state when the desired speed distribution function 
corresponds to the gamma distribution describing aggressive drivers. Both, the one and two-classes distribution functions are 
obtained. Section \ref{fds} is devoted to the construction of the fundamental diagrams for one and two classes, taking one 
class as a slow one and the second class as the fast one. Section \ref{Results} presents the numerical results obtained and the 
requirements they satisfied are explicitly shown. Lastly, in Section \ref{remarks} we give some concluding remarks.

\section{\label{PHB} The homogeneous-steady state according to the Prigogine-Herman-Boltzmann equation}
In order to set the problem, let us start with a one class of vehicles traffic flow along a highway without in/out ramps. The 
vehicles travel in the $x-$direction with speed $c$ and will be described by the distribution function $f_i(x,c,t)dxdc$ which is 
the number of class$-i$ vehicles around $(x,~c)$ at time $t$. Then, according to the traffic flow equation proposed by Prigogine 
and Herman following the steps given by Boltzmann ideas, which will be called as PHB equation, we write 
\begin{equation} 
\frac{\partial f_i}{\partial t}+c\frac{\partial f_i}{\partial x}=-\frac{f_i-f_i^{0}}{\tau_i}+\sum_j\mathcal{I}_{ij}\label{PHBe}
\end{equation}
from now and on we will write $f_i=f_i(x,c,t)$ and $f_i^0=f_i^0(x,c,t)$ if necessary to short notation but
all the distribution functions depend on $(x,t)$ unless it is specified, 
$f_0(x,c,t)$ is the desired distribution 
function and $\tau_i(\rho)$ is the relaxation time taken by the actual distribution function $f_i(x,c,t)$ to reach 
the desired $f_i^0(x,c,t)$. On the other hand, the quantity $\mathcal{I}_{ij}$ is the interaction term given by
 \begin{eqnarray}\nonumber
\mathcal{I}_{ij}&=&f_j\left(x,c,t\right)(1-p_i)\int_{w>c}(w-c)f_i(x,w,t)dw\\* \label{colision}
                &-&f_i\left(x,c,t\right)(1-p_i)
\int_{w<c}(c-w)f_j(x,w,t)dw,
\end{eqnarray}
The overtaking probability $p_i(\rho)$ plays a role and, it is a function of an effective density which will be specified 
afterwards. The macroscopic variables in this problem are given through the moments of the distribution function which are written 
as follows
\begin{eqnarray}\label{densidad}
\rho_i(x,t)&=&\int_0^{\infty}f_i(x,c,t)dc,\\
\label{velocidad}\rho_i(x,t)v_i(x,t)&=&\int_0^{\infty}cf_i(x,c,t)dc,
\end{eqnarray}
they represent the density and the average speed. %and the speed variance which gives the traffic pressure. 
Also, we must define the average desired speed given with the desired distribution function
\begin{equation}\label{v0i}
\rho_i\left(x,t\right)v_i^{0}(x,t)=\int_0^{\infty}cf_i^{0}(x,c,t)dc.
\end{equation}
Let us consider the interaction terms in the kinetic equation (\ref{PHBe}) and define a passive
\begin{equation}\label{pasivo}
 \xi_i(x,c,t)= \int_{w>c}\frac{f_i(x,w,t)}{\rho_i(x,t)}dw,
\end{equation}
and an active interaction rate
\begin{equation}\label{activo}
\psi_i(x,c,t)=\int_{w<c}\frac{f_i(x,w,t)}{\rho_i(x,t)}dw,
\end{equation}
which are related by
\begin{equation}
\psi_i(x,c,t)=(v_i-c)-\xi_i(x,c,t).
\end{equation}

Now, the interaction term is rewritten as
\begin{equation}\label{inter-a}
\mathcal{I}_{ij}=f_j(1-p_i)\rho_i\xi_i+f_i(1-p_i)\rho_j\psi_j.
\end{equation}
To go forward, let us first consider two classes: slow (class-1) and fast (class-2) drivers. Each class-2 driver 
always drives faster that any class-1 driver, implying that both the probability densities 
$\left(f_i(x,c,t)/\rho_i(x,t), ~i=1,2\right)$ have disjoint support. Let us note that the interaction terms 
$\mathcal{I}_{11}$, $\mathcal {I}_{22}$ can be immediately written as
\begin{eqnarray}\label{11-22}
\mathcal{I}_{11}=(1-p_1)\rho_1(v_1-c)f_1(x,c,t),\\
\mathcal{I}_{22}=(1-p_2)\rho_2(v_2-c)f_2(x,c,t).
\end{eqnarray}
On the other hand the interaction terms $\mathcal{I}_{12}$, $\mathcal{I}_{21}$ must be integrated taking into account the 
corresponding supports. We first consider the contributions of interactions which the fast user class-2 yields with respect 
to the kinetic equation of the user class-1. In this case the active rate
\begin{equation}\label{psi2}
\psi_2(x,c,t)=\int_{w<c}(w-c)\frac{f_2(x,w,t)}{\rho_2}dw=0,
\end{equation}
because the integration is over the class-1 support where $f_2(x,w,t)/\rho_2$ vanishes, also 
\begin{equation}\label{xi1} 
\xi_1(x,c,t)=\int_{w>c}(w-c)\frac{f_1(x,w,t)}{\rho_1}dw=0,
\end{equation}
due to the fact that the integration interval $(w>c)$ is in the class-2 support and any slow driver can go faster than 
any class-2 vehicle. It means that
\begin{equation}\label{I12}
\mathcal{I}_{12}(x,c,t)=0.
\end{equation}

Now we come back to the interaction contributions which the slow class-1 yields with respect to the kinetic equation of the user-class 2. The active interaction rate reads 
\begin{equation}\label{psi1}
\psi_1(x,c,t)=\int_{w<c}(w-c)\frac{f_1(x,w,t)}{\rho_1}dw,
\end{equation}
here the speed $c>w$ because we are referring to the faster class-2, besides when $c<w$ the probability density $f_1(x,w,t)/\rho_1$ 
vanishes and the integration interval can be extended from zero to infinity. Then, $\psi_1(x,c,t)=v_1-c$. Similarly we obtain that
$\xi_2(x,c,t)=v_2-c$, and the interaction term $\mathcal{I}_{21}$ is written as follows
%%%%%%%%%%%%%%%%%%%%%%
\begin{equation}\label{I21}
\mathcal{I}_{21}=(1-p_2)\left[\rho_2(v_2-c)f_1+\rho_1(v_1-c)f_2\right].
\end{equation}
%%%%%%%%%%%%%%%%%%%%%%%%%%%%%%%%%%%%%%%%%%%
This means that class-1 drivers do not interact with class-2 whereas class-2 feels the presence of slower vehicles which impede 
their free motion, a situation which implies some interesting aspects of the problem. Finally the model reads
\begin{eqnarray}\label{m1}
 \frac{\partial f_1}{\partial t}+c\frac{\partial f_1}{\partial x}+\frac{f_1-f_1^0}{\tau_1}&=&(1-p_1)\rho_1 \left(v_1-c\right) f_1,\\
\label{m2}
 \frac{\partial f_2}{\partial t}+c\frac{\partial f_2}{\partial x}+\frac{f_2-f_2^0}{\tau_2}&=&(1-p_2)\left[\rho_2 \left(v_1-c\right) f_1
 +\rho_1 \left(v_1-c\right) f_2-\rho_2 \left(v_2-c\right) f_2\right].
\end{eqnarray} 
To start with the analysis we will construct the homogeneous-steady distribution function, usually called as equilibrium state 
which will be obtained from the PHB kinetic equation when the drift term vanishes. Since there is neither the dependence on the 
position coordinate nor the time the kinetic equation can be solved when the desired distribution function is given. It can be 
selected according to the empirical data or be taken from the aggressive drivers model among others.
To begin with this program let us consider first the class-1 drivers, then Eq. (\ref{m1}) applied to this class can be solved, so
%%%%%%%%%%%%%%%%%%%%%%%%%
\begin{equation}\label{f1e}
f_1\left(c\right)=\frac{f_1^{0}(c)}{1-\rho_1\tau_1
\left(1-p_1\right)\left(v_1-c\right)},
\end{equation}
%%%%%%%%%%%%%%%%%%%%%%%%%%%%
where the quantities $(\tau,~p,~v)$ depend only on the density $\rho_1$. Now, $f_i(c)$ means that the distribution function
depends only on $c$. According to Prigogine \cite{Prigogine71}, one can assume that
the probability of passing and the bumper to bumper density are related by
%%%%%%%%%%%%%%%%%%%%%%%%%%%%%%
\begin{equation}\label{defs1}
 p_1\left(\rho_1\right)=1-\rho_1\ell_1 \qquad \textrm{and} \qquad \tau_1=\tau_{0}
 \frac{\left(1-p_1\left(\rho_1\right)\right)}{p_1\left(\rho_1\right)} 
\end{equation}
with $\tau_0$ constant, so we can rewrite Eq. (\ref{f1e}) as 
%%%%%%%%%%%%%%%%%%%%%%%%
\begin{equation}\label{f1e-a}
f_1(c)=\frac{f_1^{0}(c)}{a_1+\gamma_1 c},
\end{equation}
%%%%%%%%%%%%%%%%%%
where $a_1=1-\gamma_1 v_1$ and 
\begin{equation}\label{defs} 
\gamma_1=\frac{\tau_0}{\ell_1}\frac{ z_1^{3}}{\left(1-z_1\right)},
\end{equation}
with $z_1=\rho_1\ell_1$, here $\ell_1$ is the size of class-1 vehicles. Note that $\gamma_1$ depends on $\rho_1$ whereas 
$a_1$ depends on the slow-class density and the mean speed $v_1$.

The homogeneous and steady distribution function given in Eq. (\ref{f1e-a}) must satisfy
the following two requirements imposed in Eqs. (\ref{densidad}) and (\ref{velocidad})
\begin{eqnarray}\label{ro1}
\rho_1=\int_{0}^{\infty}f_1\left(c\right)dc,\\
\label{q1}
\mathcal{Q}_1=\rho_1v_1=\int_0^{\infty}cf_1\left(c\right)dc.
\end{eqnarray}
The fast-class homogeneous-steady state distribution function  can be obtained from Eq. (\ref{m2}) yielding
\begin{equation}
  f_2\left(c\right)=\frac{f_2^0\left(c\right)}{1-\tau_2\left(1-p_2\right)\left(\mathcal{Q}-\rho c\right)} 
  +\frac{\rho_{2}\tau_{2}\left(1-p_{2}\right)\left(v_{2}-c\right)f_1^0\left(c\right)}{\left(1-\tau_2(1-p_2)(\mathcal{Q}-\rho c)\right)
  \left(1-\tau_1(1-p_1)(\mathcal{Q}_1-\rho_1 c)\right)},\label{f2e}
\end{equation}
where $\rho=\rho_1+\rho_2$ is the total density and ${\mathcal Q}=\rho v=\rho_1v_1+\rho_2 v_2={\mathcal Q}_1+{\mathcal Q}_2$ 
the total flux. Now due to the fact that class-2 is coupled with the slow class-1, 
the probability of overpassing $p_2$ will be affected by the total density instead of the partial density $\rho_2$ only. 
In fact, the effective density seen by class-1 is the $\rho_1$ density, whereas the one seen by class-2 is total density 
$\rho$, then
\begin{equation}\label{p2}
p_2=1-\rho l,\qquad \tau_2=\tau_0\frac{(1-p_2)}{p_2},
\end{equation}
and we define
\begin{equation}
l=\frac{\left(\rho_1 \ell_1+\rho_2 \ell_2\right)}{\rho}, \label{L}
\end{equation}
note that the quantity $l$ is a kind of effective vehicular length that depends on vehicular concentration. Now, one can rewrite Eq.(\ref{f2e}) as
\begin{equation}
 f_{2}\left(c\right)=\frac{f_{2}^{0}\left(c\right)}{a+\gamma c}+\frac{\tau_0\left(1-\sigma\right)z^2}{\left(1-z\right)}
 \frac{\left(v_{2}-c\right)f_{1}^{0}\left(c\right)}{\left(a+\gamma c\right)
 \left(a_{1}+\gamma_{1}c\right)}\label{f2e-2}
\end{equation}
here we have introduced $a=1-\gamma v$ and
\begin{equation}
 \gamma=\frac{\tau_0}{l}\frac{z^3}{\left(1-z\right)},
\end{equation}
where $z=\rho l$ and $\sigma=\rho_1/\rho$ is the concentration of slow drivers.
Again, the equilibrium distribution function given in Eq. (\ref{f2e-2}) must satisfy
the following two requirements imposed in Eqs. (\ref{densidad}) and (\ref{velocidad})
\begin{eqnarray}\label{rho2}
\rho_2=\int_{0}^{\infty}f_2\left(c\right)dc,\\
\label{q2}
\mathcal{Q}_2=\rho_2v_2=\int_0^{\infty}cf_2\left(c\right)dc.
\end{eqnarray}
All these quantities can be calculated once we have chosen the desired distribution function. 

\section{The fundamental diagram }\label{fds}
\subsection{The slow user-class}
Now, let us consider the gamma distribution function, which describes the behavior of 
aggressive drivers through a parameter $\alpha $ quantifying their  acceleration. It is given as  
\begin{equation}\label{gamma} 
f_i^{0}\left(c\right)=\frac{\rho_i}{v_i^{0}}\frac{\alpha}
{\Gamma\left(\alpha\right)}\left(\frac{\alpha c}{v_i^{0}}\right)^{\alpha-1}\exp\left(-\frac{\alpha c}{v_i^{0}}
\right),
\end{equation}
where $\alpha>1$ is the shape parameter. The gamma distribution function given in Eq. (\ref{gamma}) is introduced in equation 
(\ref{f1e-a}), which must satisfy the condition (\ref{ro1}) yielding 
\begin{equation}
 b_1=\exp\left(\mu_1\right)E_{\alpha}\left(\mu_1\right),\label{norma1}
\end{equation}
where $b_1=\gamma_1/\alpha$, $\mu_1=a_1/b_1$ and $E_n\left(x\right)$ is the exponential integral function.
Recalling that $0\leq\gamma_1<\infty$, when $\gamma_1=0$ the quantity $a_1=1$ and as a 
consequence an increase on $\gamma_1$ means a decrease on $a_1$, which in fact can not be negative. 
This fact allows us to 
define a critical value $\gamma_1^c$ when $a_1=0$, and also a critical reduced density $z_1^c$ given through
\begin{equation}\label{gammac}
\beta_1\left(z_1^c\right)^3-\frac{\alpha}{\alpha-1}\left(1-z_1^c\right)=0,
\end{equation}
where $\beta_1=v_1^0\tau_0/\ell_1$. In the collective regime (i.e., above the reduced critical density) the traffic flow of
the slow user-class $\mathcal{Q}_1$ is characterized by the function
\begin{equation}
 \mathcal{Q}_1=\frac{1}{\tau_0}\frac{\left(1-z_1\right)}{z_1^2}, \quad z_1>z_c
\end{equation}
which is independent of the shape parameter $\alpha$. In the individual regime (i.e., below the reduced critical density) the traffic flow
of the slow user class yields
\begin{equation}
 \mathcal{Q}_1=\frac{\left(1-a_1\right)}{\tau_0}\frac{\left(1-z_1\right)}{z_1^2}, \quad z_1<z_c \label{Q1}
\end{equation}
where $a_1$ must be determined numerically from Eq. (\ref{norma1}).

\subsection{The fast user-class}
For the fast user class we proceed analogously as in the former case. The distribution function in Eq. (\ref{f2e-2}) 
must determine the fast class density, when we insert Eq. (\ref{gamma}) in (\ref{rho2}) a straightforward calculation
yields
\begin{eqnarray}\nonumber
 b=\exp\left(\mu\right)E_{\alpha}\left[\mu\right]&-&b\sigma\frac{\mu_{1}}{\left(\mu_{1}-\mu\varepsilon\right)}
 \left[1-\frac{\mu\varepsilon}{\mu_{1}}\frac{\exp\left(\mu\varepsilon\right)E_{\alpha}
 \left[\mu\varepsilon\right]}{\exp\left(\mu_{1}\right)E_{\alpha}\left[\mu_{1}\right]}\right]\\
&+&\frac{\sigma}{1-\sigma}\frac{\varepsilon}{\left(\mu_{1}-\mu\varepsilon\right)}
\left(1-\mu b-\tau_{0}\mathcal{Q}_{1}\frac{z^{2}}{1-z}\right)\left[\frac{\exp\left(\mu\varepsilon\right)E_{\alpha}
\left[\mu\varepsilon\right]}{\exp\left(\mu_{1}\right)E_{\alpha}\left[\mu_{1}\right]}-1\right],\label{norma2}
\end{eqnarray}
where $b=\gamma v_2^0/\alpha$, $\mu=a/b$ and $\varepsilon=v_2^0/v_1^0$. Equations (\ref{norma1}) and (\ref{norma2}) 
must be solved simultaneously for $\mu_1$ and $\mu$ in order to obtain the relation $\mathcal{Q}(\rho)$. The solution was
found numerically and the results are shown in Figure \ref{fig1}. 
The critical reduced density $z_c$ is established by setting $a=0$ in Eq. (\ref{norma2}), which yields
\begin{equation}
 z_{c}^{3}\left(\left(1-\sigma^{2}\right)\beta+\sigma\varepsilon\beta_{1}\right)-\frac{\alpha}{\alpha-1}
 \left(\left(1-\sigma\right)+\sigma\varepsilon\right)\left(1-z_{c}\right)=0,\label{zc}
\end{equation}
where $\beta=v_2^0\tau_0/l$, note that in the case $\sigma=0$ we recover Eq. (\ref{gammac}). In the individual regime the total traffic flow is given
by
\begin{equation}
\mathcal{Q}=\frac{\left(1-a\right)}{\tau_{0}}\frac{\left(1-z\right)}{z^{2}},\label{Q}
\end{equation}
with $a\neq 0$ while the collective traffic flow is obtained by setting $a=0$ in Eq.(\ref{Q}).

\section{Numerical results and comparison with empirical data}\label{Results}
\begin{figure}
\includegraphics[scale=0.68]{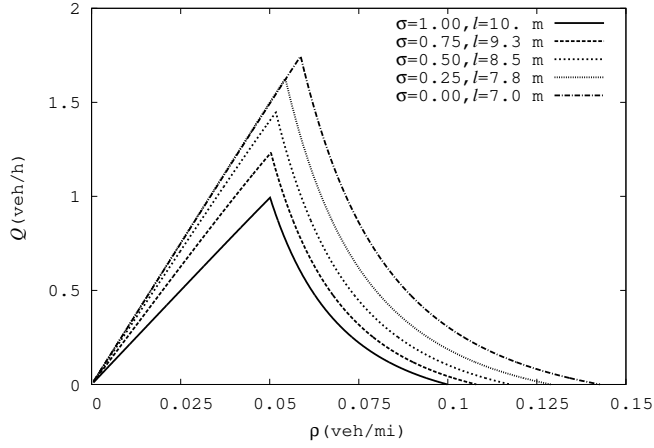}
\caption{The total flux in $Q\left[\textrm{veh/s}\right]$ versus the total density $\rho\left[\textrm{veh/m}\right]$ 
for different values of the density fraction $\sigma$ corresponding to different effective lengths $l$ for 
$\alpha=120$, $\varepsilon=3/2$ and $\tau_0=2~s$.}
\label{fig1}
\end{figure}

The flux-density relation for the two-class mixture of drivers can be obtained by solving simultaneously Eqs. (\ref{norma1}) 
and (\ref{norma2}) for $\mu_1$ and $\mu$, from those quantities one can obtain $a_1$ and $a$ to finally determine the fundamental
diagram from Eqs. (\ref{Q1}) and (\ref{Q}). In Ref. \cite{Coifman15}, Coifman observed that many of the parameters of the 
flow-density relationship depend on vehicle length, moreover there is obtained a fundamental diagram for each class of vehicle, 
classified by its size. In Figure \ref{fig1} we show traffic flow as a function of the density for 
different values of the fraction $\sigma$, or equivalently different values of the effective length $l$, for $\alpha=120$, 
$\varepsilon=3/2$ and $\tau_0=2~s$. Note that when fixing the lengths 
$\ell_1$ and $\ell_2$, for the slow and fast vehicles respectively, an effective length $l$ is established for every given value 
of $\sigma$ (see relation (\ref{L})). In Figure \ref{fig1} one can appreciate the separation of the two regimes, induvidual and 
collective, by the critical density $\rho_c$. Note that, accordingly with data presented in Ref. \cite{Coifman15} the critical 
density moves to the left in the $\rho-\mathcal{Q}$ plane as the effective length $l$ increases and the road capacity 
decreases as $l$ increases.

\begin{figure}
\centering
\includegraphics[scale=0.68]{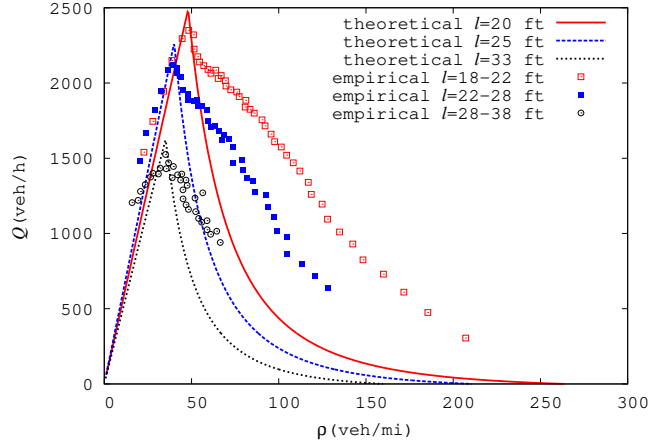}
\caption{The flow-density relation: the lines represent the theoretical predictions derived
from the non-local Prigogine-Herman kinetic equation for a gamma distribution, while the dots
represent the traffic data measured by Coifman \cite{Coifman15}. We have used the following parameters: 
%Top: $\alpha=100$, $\tau_0=54~s$,$\ell_1=28~ft$, $\ell_2=18~ft$ and $\epsilon \approx 9/7$. The correspondence between $\sigma-l$ is as follows: 
%$\sigma=1.0$ - $l=28 ft$, $\sigma\approx 0.38$ - $l=22 ft$ and $\sigma=0.0$ - $l=18 ft$. 
$\alpha=120$, $\tau_0=35~s$, $\ell_1=33~ft$, $\ell_2=20~ft$ and $\epsilon \approx 1.12$. The correspondence between $\sigma-l$ 
is as follows: ($1.0~$-$~33 ft$), ($0.38~$-$~25 ft$) and ($0.0~$-$~20 ft$).}\label{fig2}
\end{figure}

In figure \ref{fig2} we show the comparison between our theoretical model with Coifman empirical data for different effective 
lengths. In fact we consider a mixture where we have set $\sigma=0$ for $l=20~ft$ and $\sigma=1$ for 
$l=33~ft$, then through the 
relation (\ref{L}) one can obtain a particular value of $l$ from every given $\sigma$, ($0 \leq \sigma \leq 1$). For short 
vehicles the theoretical $Q-\rho$ relation fits well experimental data in the individual regime. For example the 
critical density $\rho_c$, the traffic capacity $Q_0$ and $v_{max}$ are well described. In the collective region and for large 
vehicles some other factors, as space requirements of vehicles must be considered, this generalization would be accomplished 
elsewhere. 

\begin{figure}
\centering
\includegraphics[scale=0.45]{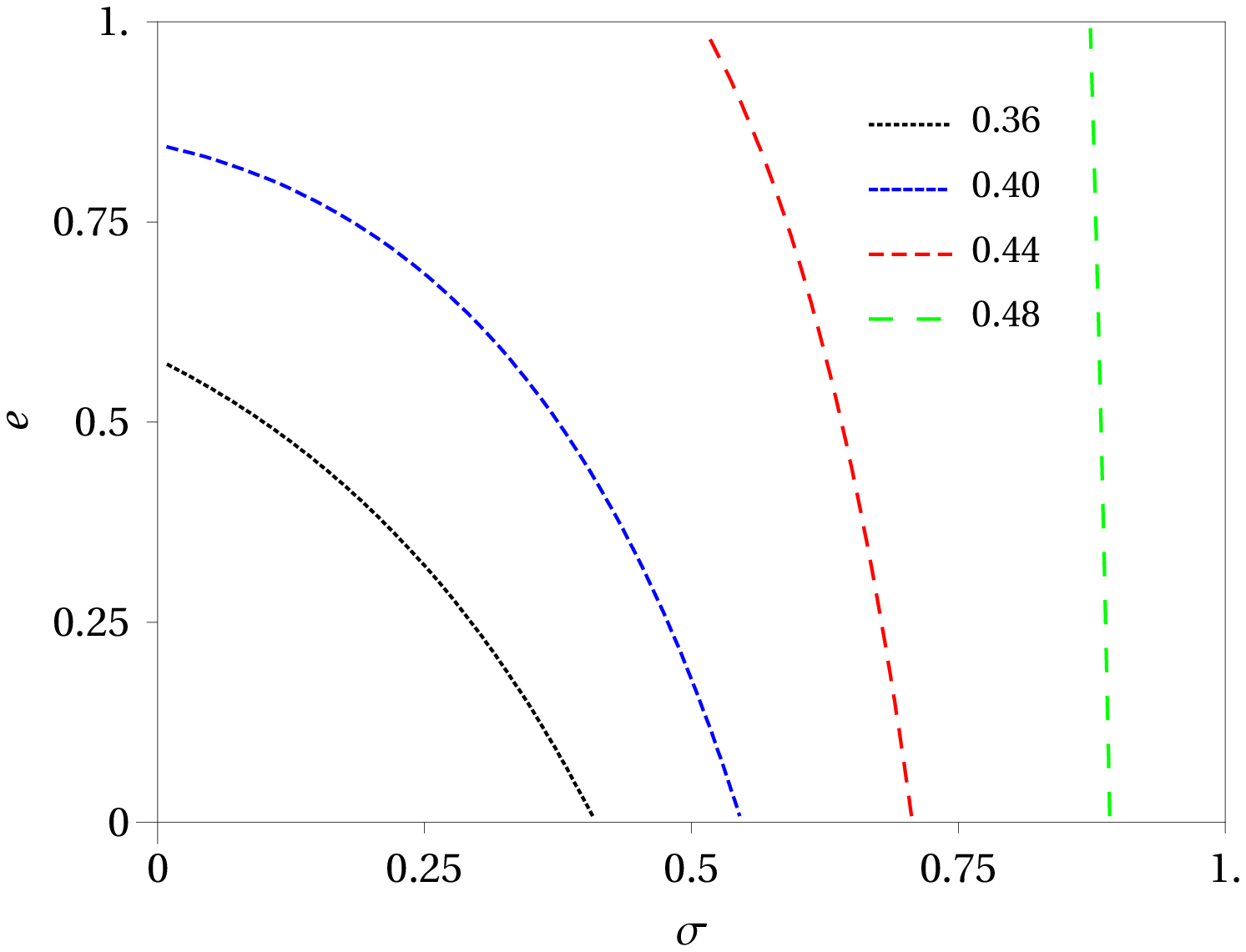}
\includegraphics[scale=0.45]{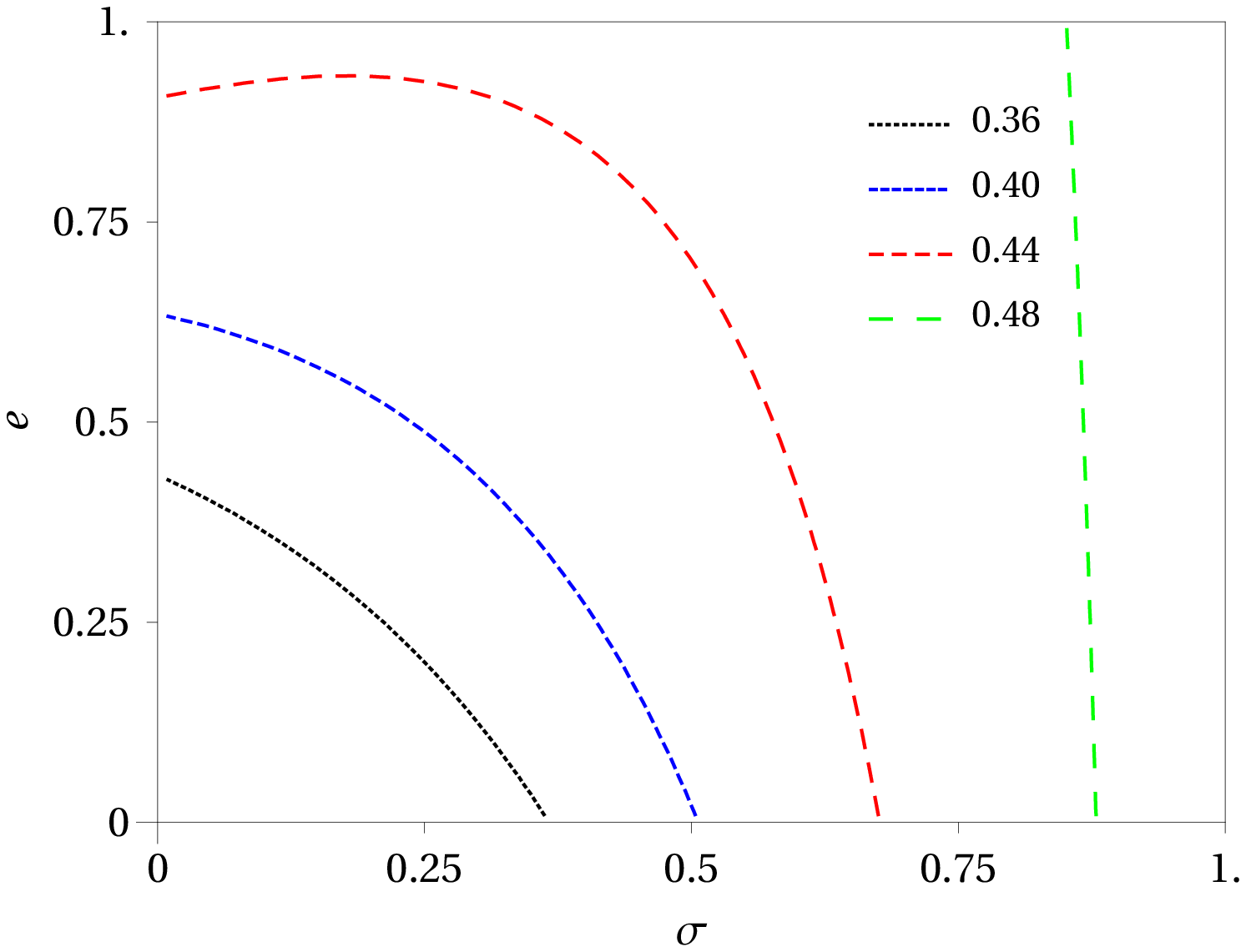}
\caption{Contour levels at constant $z_c$ as a function of $\sigma$ and $e=\ell_1/\ell_2$ and for two different values
$\varepsilon=2$ (left) and $\varepsilon=3/2$ (right).}
\label{fig3}
\end{figure}

At figure \ref{fig3} the dependence of $z_c$ with $\sigma$ and $e=\ell_2/\ell_1$ for two different values of $\epsilon$. 
The parameter $e$ is a measure of the disparity of the lengths $\ell_i  ~(i=1,2)$, $e$ approaches to unity as 
$\ell_2\rightarrow \ell_1$ and approaches to zero as $\ell_2 \rightarrow 0$ for $\ell_1$ fixed. In general, the critical 
value $z_c$ may remain fixed as $e$ increases and $\sigma$ decreases.

\section{Concluding Remarks}\label{remarks}
In this work, a theoretical model for multiclass traffic based on the Prigogine-Herman-Boltzmann equation \cite{Prigogine71} 
has been determined. It has been shown that the fundamental diagram, as well as other important quantities such as the critical 
density and the road capacity can be derived from a theoretical model. The flux-density relation has been solved numerically 
from our model, and the obtained diagram matches experimental data collected by Coifman in the individual regime for different 
values of the effective length $l$. However, we have noted that at moderate densities vehicles cannot be regarded as point-like 
objects, in fact they require a finite space for its own extension and reaction. In this sense it is necessary to incorporate 
some modifications to the PHB traffic equation, this aspects will be addressed in a near future.
\linebreak

\subsection*{acknowledgments}
The authors acknowledge support from CONACyT through grant number CB2015/251273.

\bibliography{PHBfund-biblio}
\bibliographystyle{unsrt}

\end{document}